%% file: sample-sigconf.tex
\documentclass[11pt]{article}
\usepackage[letterpaper,margin=1in]{geometry}
\usepackage{booktabs} 
\usepackage{textcomp}
\usepackage{gensymb}
\usepackage{graphicx}
\usepackage[square,sort,comma,numbers]{natbib}

\begin{document}

\title{The Identification and Analysis of Indicators for Predicting Malarial Incidence in Zimbabwe%
\thanks{Distribution of this paper is permitted under the terms of the Creative Commons license CC-by-nc-nd 4.0.}
}

\author{Booma Sowkarthiga Balasubramani\textsuperscript{1}, Marco Nanni\textsuperscript{1,2}, Shin Imai\textsuperscript{1}, Isabel F. Cruz\textsuperscript{1} \\[12pt]
\textsuperscript{1}University of Illinois at Chicago, USA
\textsuperscript{2}Politecnico di Milano, Italy
\\
\textsf{\{bbalas3,mnanni2,simai2,ifcruz\}@uic.edu}
}


\maketitle

\begin{abstract}
With over 50\% of the country's population at risk of contracting malaria despite the introduction of several measures to combat the disease, Zimbabwe is one of the eight countries in the Malaria Elimination 8 platform of the Southern African Development Community. Various indicators, including temperature, population distribution, land cover, and access to hospitals affect the incidence and spread of this disease. In this paper, we consider different such indicators and present our analysis of their interaction (e.g., how the Plasmodium falciparum Parasite Rate (PfPR) affects the sickle cell trait) and their effect on malaria incidence in Zimbabwe. We also discuss the results of our preliminary experiments on predictive analytics of malaria incidence based on the indicators we have considered.
\end{abstract}
%
%




\input{samplebody-conf}

\bibliographystyle{abbrv}

\bibliography{sample-bibliography} 

\end{document}

%% file: samplebody-conf.tex
\section{Introduction}

According to the recent World Health Organization (WHO) estimates, Zimbabwe is one of the eight countries in the Malaria Elimination 8 platform of the Southern African Development Community. Of the country's 63 districts, 47 districts are malarial, with 33 of them categorized as high burden malaria areas. There are more than 400,000 malaria cases each year in Zimbabwe, which translates to 3\% of the country's total population contracting the disease.  

Surveillance for malaria elimination enables monitoring the changing disease patterns based on various indicators. Several studies~\cite{abellana2008spatio,beck2013effect,ferrao2016spatio,darkoh2017} have been conducted in the past focusing
on different African countries, all with the same goal of identifying
indicators that could possibly influence malaria incidence in
the population at only one spatial resolution (country, province, or district level).
Since temperature and precipitation influence the lifecycle of Anopheles
mosquitoes, which are responsible for malaria transmission, they are the two most widely studied malarial indicators. 

Factors such as temperature, precipitation, land cover (e.g., hills, rivers, forests), vector breeding places, population distribution, and location of health facilities also affect the number of malaria cases and deaths. Including these indicators in addition to temperature and precipitation, can improve the accuracy of predicting malaria incidence at different spatio-temporal resolutions. Therefore, it is necessary to combine these data to make it possible to predict where the disease incidence is most likely to increase or decrease. However, the availability of data is not sufficient per se to bring the concept of data integration to its full potential since data come from disparate sources, are fundamentally heterogeneous, lack metadata, and have different spatial and temporal resolutions.
This state of affairs motivates the need for data transformation and data cleaning. Given the huge volume of data available and the complexity of integrating heterogeneous spatial data, the most crucial task is the determination of which indicators need to be considered for data integration, so as to obtain precise results in the prediction process.

Recent work uses one to few indicators to predict malaria in different regions of Africa~\cite{abellana2008spatio,ferrao2016spatio,zacarias2011spatial} without addressing the fact that spatial data are heterogeneous. Therefore, the main difference between our work and that of others is that we apply several indicators at different spatial resolutions (district, province and country level) to predict malarial incidence. In this paper, we discuss various indicators that affect the incidence and spread of malaria,  study the selection of indicators among highly heterogeneous data to facilitate the analysis of trends in malaria patterns, and report on the results of our preliminary experiments.

\section{Identification and Mapping of Malarial Indicators}
Decision making applications driven by big data, such as predictive analytics for malaria incidence, require the analysis of various attributes in different datasets derived from diverse open data sources, many of which are in a non-standardized format. For example, the population distribution of Zimbabwe is available as a CSV file, as opposed to the land cover dataset, which is in a standardized shapefile format). Standardization of datasets and coordination among data providers, such as healthcare facilities and government agencies, can improve the accuracy of the data being analyzed. Mapping denotes the process of converting data from a non-standardized format into a standardized format. This is accomplished by translating the data obtained from heterogeneous sources into a common spatial data format (e.g., shapefile) using QGIS~\cite{qgis}. The datasets that have been identified thus far for potential inclusion in the malaria incidence prediction application, their  data formats, and their spatio-temporal resolutions are described as follows:
\begin{description}
\item[Plasmodium falciparum Parasite Rate (PfPR)] \emph{Plasmodium falciparum} is a parasite that is known to be a common cause of malaria in humans. PfPR (number of malaria cases per 1000 people) data for Zimbabwe is obtained as a TIFF file from a non-proprietary database, with a country level resolution. 
The {\tt gdal\_translate} function in {\tt OSGeo4w} is used to extract these data for the years from 2000 to 2015. 

\item[Land cover] Land cover data for Zimbabwe is extracted from the {\tt RCMRD geoportal} ~\cite{rcmrd} as a TIFF file, and the {\tt gdal\_translate} function in {\tt OSGeo4w} is used to convert it to a standardized format at a country level spatial resolution. 
\item[Hospital locations] The data on the list of hospitals in Zimbabwe are available at the district and province level spatial resolution, and the Geonames {\tt geocoder} web service was used to associate hospital names to latitude and longitude.
\item[Elevation] These data is obtained for each of the 63 districts in Zimbabwe, from which the average elevation for each province is computed. 

\item[Weather] Data on temperature (\degree C) and rainfall (mm) for each month at the country level spatial resolution for the years 1991 to 2015, allowed the understanding of differences in dry and wet seasons in Zimbabwe, and is expected to be highly correlated to malaria transmissions. Monthly temperature, wind speed and humidity data for the year 2015 is obtained at the district level from Weather History \& Data Archive for Gweru and Chipinge, districts with the low and high rates of malaria cases and deaths respectively. The dataset from the National Centers for Environmental Information (NCEI) that contains data recorded from 28 weather stations in different districts of Zimbabwe for 2015, the same year for which the number of malaria cases and deaths data for each district is available, is also considered for our study. This dataset required identification of the station numbers for the corresponding districts, and integration of the data for two or more stations of the same district. 

\item[Population and malaria incidence] Data on the number of cases and deaths for the years from 1991 to 2015 is obtained from the population dataset about malaria in Zimbabwe, from which the number of cases and deaths with respect to the age, for both males and females were extracted. As indicated by various studies~\cite{kanyangarara2016individual,ferrao2016spatio,abellana2008spatio}, the number of deaths vary significantly between different age groups such as children under 1 year of age, people under 25 and over 25 years of age. Therefore, data aggregation has been performed on these data to group the population into 6 different categories based on their age: (a) children <1 year; (b) children between 1 and 4 years; (c) 5 to 24 years; (d) 25 to 49 years; (e) 50 to 69 years; and (f) 70 plus years. The latest census data (for the year 2012) for Zimbabwe which provides insights on the rate of malaria cases and deaths at the district and province level resolutions is also considered.
\item[Malaria cases and deaths] This province and district level data was extracted from the 2017 President's Malaria Initiative file, for the year 2015 with information on the number of health facility confirmed malaria cases, number of village health worker confirmed malaria cases, and the total number of malaria deaths.  
\item[Sickle cell frequency] \emph{Sickle cell trait} is an inherited blood disorder that  is known to cause significantly fewer deaths due to malaria, especially when Plasmodium falciparum is the causative organism. \emph{Sickle cell frequency} or \emph{sickle haemoglobin allele frequency}, which denotes the frequency of the population affected by sickle cell trait, was obtained from the Malaria Atlas Project for the year 2013 at the country level spatial resolution.
\end{description}

\section{Correlation Analysis}
Given the number of datasets and the attributes in each dataset, identifying which attributes can contribute significantly to the malaria incidence helps increase the accuracy of prediction. Therefore, it is important to determine the correlation among various attributes in different datasets. \emph{Correlation} is a statistical measure which indicates the relationship between two or more attributes, that is, the degree to which the attributes are associated with each other, such that the change in one attribute is accompanied by the change in another. The results of the correlation analysis are discussed as follows:
\begin{figure*}
  \includegraphics[width=\textwidth,height=5cm]{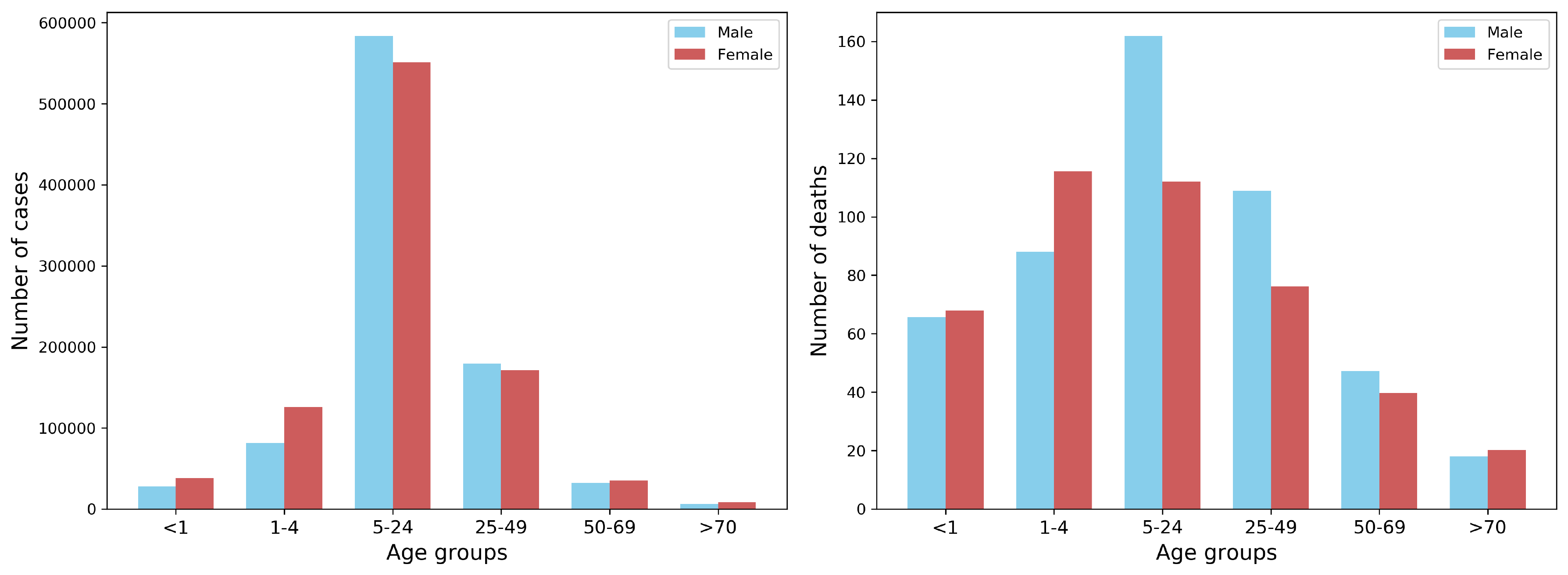}
  \vspace{-0.4cm}
  \caption{Average number of malaria cases and deaths by group and gender.}
  \label{fig1}
\end{figure*}
\paragraph{Malaria cases and deaths vs.\ Population} As reported by several studies in the past~\cite{kanyangarara2016individual,ferrao2016spatio,abellana2008spatio}, the number of malaria cases and deaths vary substantially with respect to age and gender. The number of cases and deaths for different age groups is summarized in Fig~\ref{fig1}. 
It indicates that the average rate of incidence in females is higher for children under 5 years, at the country level resolution.
Also, even if the group from 1 to 4 years has a lower incidence rate compared to age groups 5 to 24, the average death rate is higher for the former.

\begin{figure}[h]
\begin{center}
\includegraphics[width=4in]{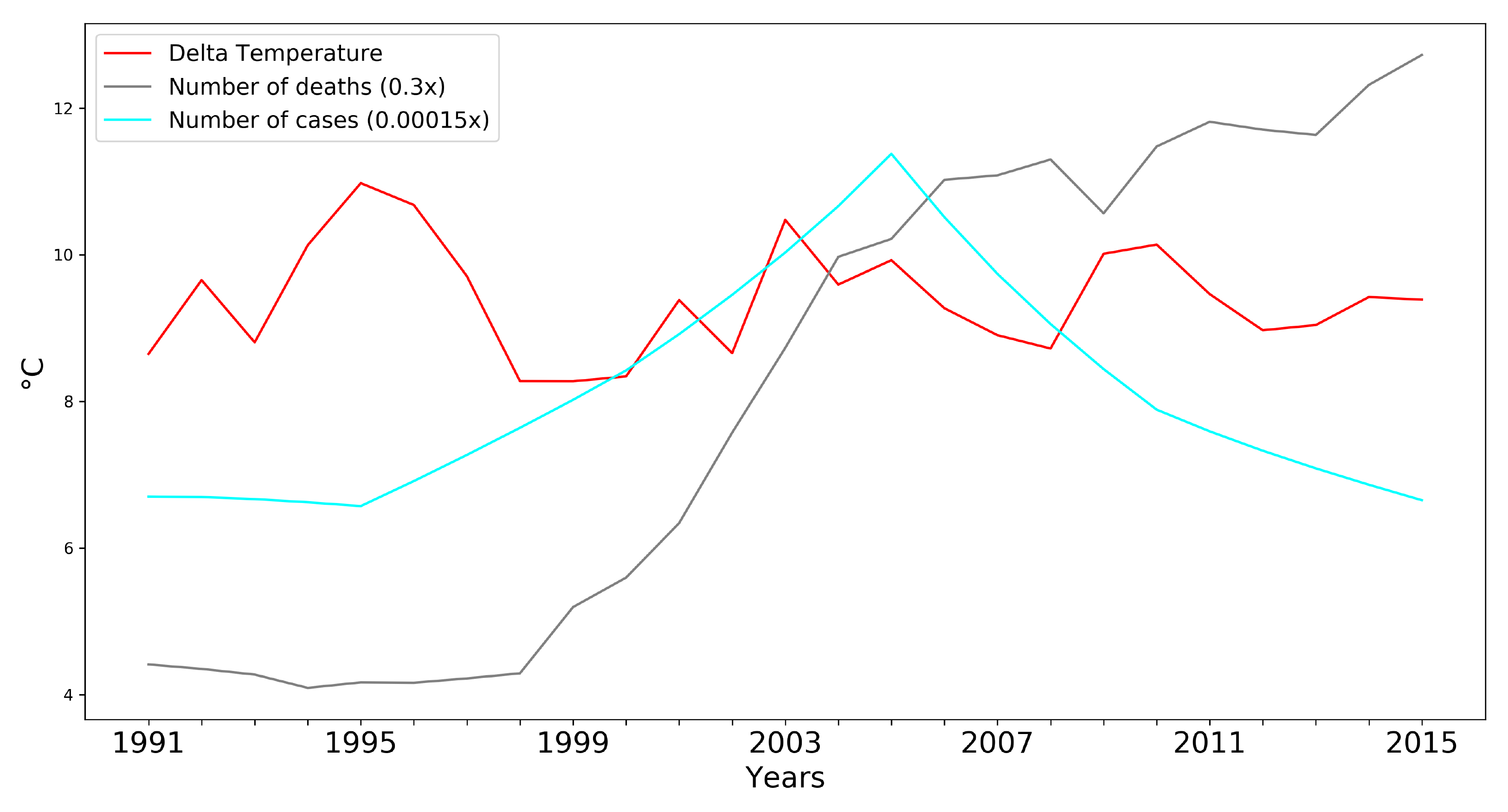}
\end{center}{}
\caption{Temperature difference vs.\ Malaria deaths/incidence over the years.}
\label{fig2}
\end{figure}

Based on the malaria cases and deaths data for each of the 63 districts of Zimbabwe, the percentage of malaria cases and deaths over each province was calculated. The results indicated that the three rural provinces like Manicaland, Mashonaland Central and Mashonaland East have approximately 83\% of all malaria cases and 64\% of all malaria deaths in 2015. It is also to be noted that the Harare region has a very low rate, despite being on the east side. This could be attributed to the fact that Harare City, the capital of Zimbabwe is very well equipped with hospitals and health care facilities, with a better road network connecting them. However, this point will be extended in the future work. The trend in the percentage of malaria deaths for each province was similar to the malaria cases, even if the number of deaths in each province is orders of magnitude smaller than the number of malaria cases. 
\paragraph{Malaria cases and deaths vs.\ Temperature} Though some studies~\cite{freeman1996temperature,mabaso2006spatio} claimed that there is a correlation between the increase in temperature and the malaria incidence and number of deaths due to malaria, our experiments with the spatial resolution at country level indicated otherwise (Fig~\ref{fig2}). 

\begin{figure*}[hbt]
\begin{center}
    
  \includegraphics[width=14cm,height=6cm]{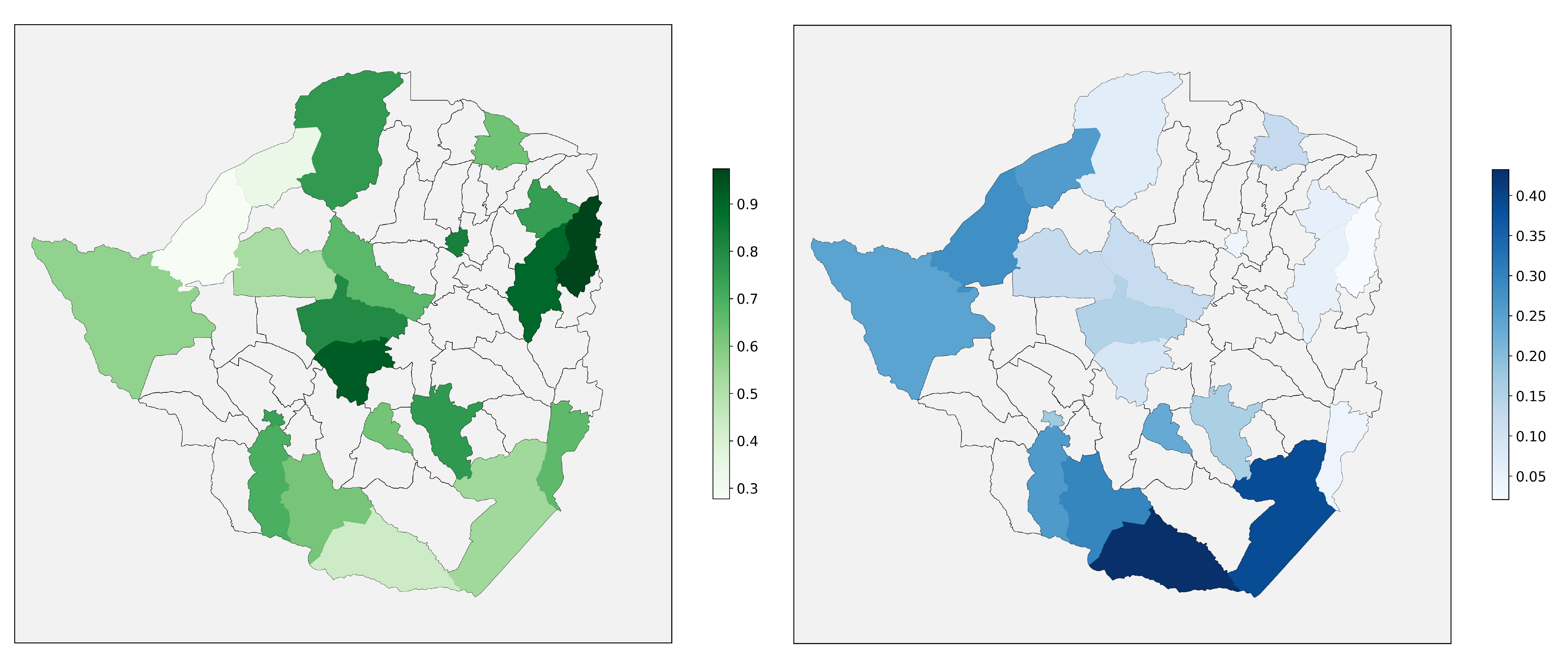}
  \end{center}
\vspace{-12pt}
  \caption{Temperature vs.\ Malaria cases: (a) Number of days with minimum temperature below 17\degree C over the total recordings; (b) Number of days with maximum temperature above 33\degree C over the total recordings.}
  \label{fig3}
\end{figure*}

According to Beck-Johnson et al.~\cite{beck2013effect}, temperature lower than 17\degree C and higher than 33\degree C is respectively too cold or too hot for both larvae and adult mosquitoes to survive, while temperature in the range 20\degree C - 30\degree C is the optimal range for mosquitoes' growth. Therefore, for each of the 28 districts and for year 2015, attributes such as the number of days with minimum temperature lower than 17\degree C (62.6\degree F), the number of days with maximum temperature higher than 33\degree C (91.4\degree F), and the number of days with minimum temperature higher than 20\degree C and maximum temperature lower than 30\degree C (68\degree F - 86\degree F) were considered, among which, the first two groups represent hard conditions for mosquitoes to reproduce and survive, while the third group is the perfect habitat for mosquitoes' lives. Fig~\ref{fig3} represents the regions with hard conditions for mosquitoes to survive and reproduce, that is, the darker the color of the region (green or blue), the tougher the climate and mosquitoes' growth (districts missing data are represented in background color). It is evident that for some regions it is true that a high number of days is correlated to a low malaria incidence, but not for other regions, which concludes that there is no significant correlation between malaria incidence and temperature, even at district level resolution. The regions in the North-east side of the country present a low number of days with maximum temperature above 33\degree C but also a high number of malaria cases and deaths. Contrarily, regions in the North and South side of Zimbabwe present more days with maximum temperature above the threshold, that is, most days in year 2015 that cause difficult conditions for mosquitoes to survive, but also a low rate of malaria incidence.

\begin{figure}[h]
\begin{center}
\includegraphics[width=5in]{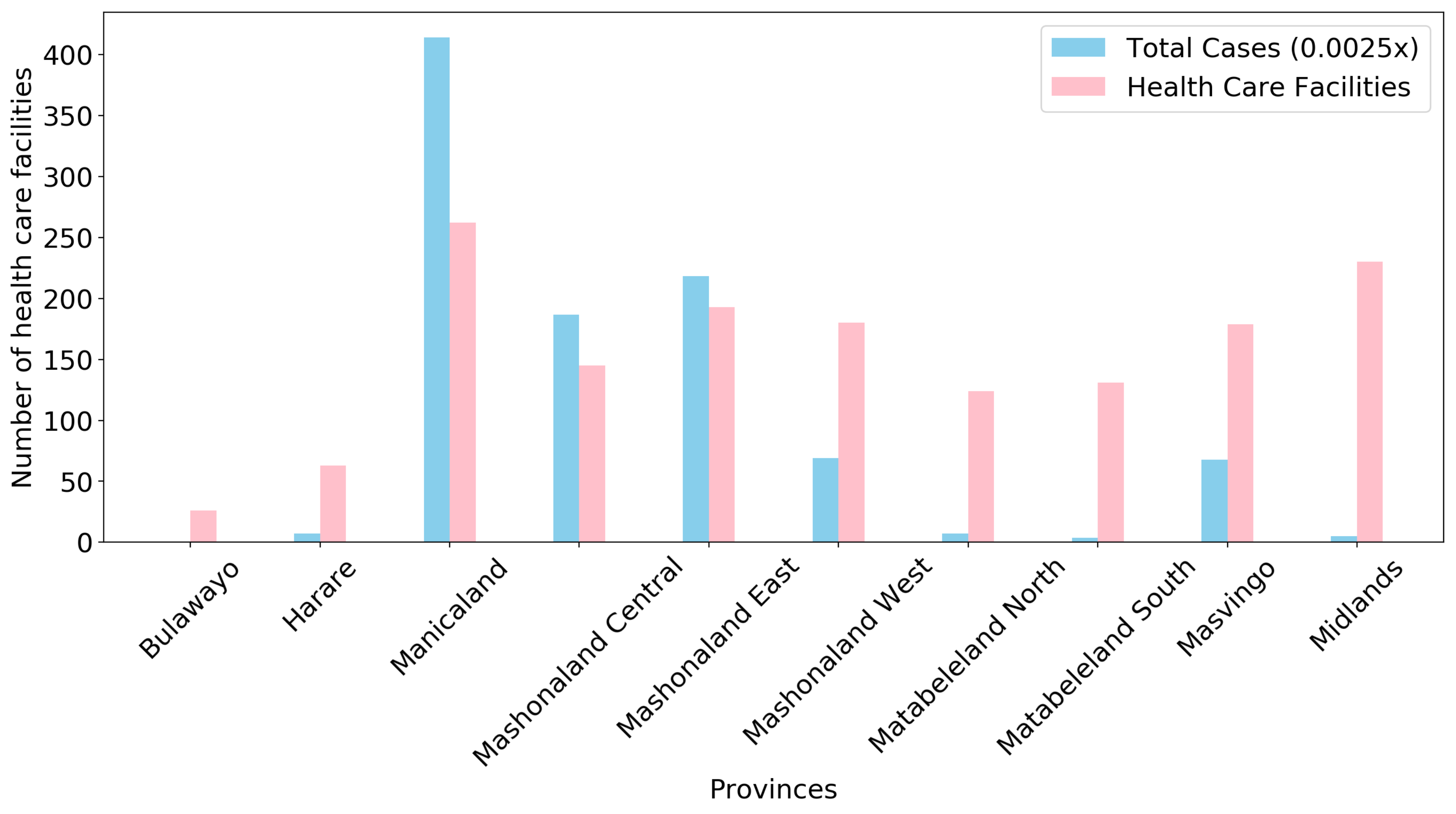}
\end{center}{}
\vspace{-0.5cm}
\caption{Malaria cases vs.\ Number of healthcare facilities.}
\vspace{-0.3cm}
\label{fig6}
\end{figure}

Minimum temperature above 20\degree C and maximum temperature below 30\degree C, correspond to the optimal conditions for mosquitoes to survive and reproduce, which ideally should be associated to a high number of malaria cases and deaths. However, though north-east Zimbabwe do not have this optimal temperature, this region has the highest number of malaria cases and deaths. To have concrete results about these considerations, \emph{Pearson linear correlation coefficient} was used to determine the correlation between malaria incidence and weather. The results showed a moderate negative linear correlation among the number of days with maximum temperature above 33\degree C (Rates Max Count) and both malaria cases and deaths, which means that an increase in the number of days with very high temperature (33\degree C) is associated with a decrease in the number of malaria cases and deaths. 
\paragraph{PfPR vs.\ Malaria cases} 
The Pearson correlation coefficient indicated that there is little association between PfPR and malaria cases or deaths. However, looking at the PfPR data for 2013, the parasite rate is particularly higher in the area of Kariba Lake in north-west Zimbabwe, while it is not so predominant in regions with the highest number of malaria cases and deaths. This led us to consider a potential indicator called sickle cell trait, which explains this situation.
\paragraph{Sickle cell trait vs.\ Malaria cases}
By considering the mean estimates of sickle cell frequency in the Zimbabwean population, it is clear that areas around Kariba Lake present a high rate of sickle cell trait, a natural selector and advantage against malaria~\cite{luzzatto2012}. These areas also present a high rate of P. falciparum parasite. Thus, in Zimbabwe sickle cell trait is found very high correlated to PfPR. This  explains why these areas do not present the highest rates of malaria cases and deaths, the reason being sickle cell trait has been over the years an advantage for people coming from those regions. The same advantage is not present for areas most affected by malaria in these recent years, for example, those in eastern Zimbabwe bordering Mozambique. These areas do not present a high frequency of sickle cell allele, neither a high parasite rate. However, Mozambique presents very high values of both PfPR and sickle cell frequency, which concludes that it has been one of the countries affected by malaria in Africa over the years, with a population that has been developing a high rate of sickle cell and a decreasing parasite rate since the year 2000. 
\paragraph{Malaria cases vs.\ Elevation} Mabaso et al.~\cite{mabaso2006spatio} suggested that altitude could be one among the factors related to malaria incidence at the district level. However, our experiments at the province level confirms that there is no correlation between elevation and malaria cases.
\paragraph{Humidity and wind vs.\ Malaria cases} The higher the percentage value of humidity during the year, the higher the number of malaria cases and deaths. Thus, as indicated by Zacarias and Andersson~\cite{zacarias2011spatial}, humidity is a potential factor that influence malaria incidence. On the other hand, there seem to exist an opposite correlation with the wind speed. That is, the lower the wind speed, the higher the number of malaria cases.
\paragraph{Malaria cases vs.\ Healthcare facilities} The Pearson correlation coefficient value for malaria total cases and health care facilities is 0.62 and for malaria total deaths and health care facilities, it is 0.54. This confirms that there is a high correlation between malaria incidence and health care facilities in each province (Fig~\ref{fig6}). This indicator will be studied in detail along with other potential indicators as part of the future work, to analyze the prevention and treatment methods for malaria.

\section{Predictive Analytics}

The lack of non-proprietary data for malaria incidence apart from the year 2015 meant that there is only limited amount of data to work with. Predictive analytics in this case was difficult because of the small amount of data that is available for the training dataset. Consequently, the dataset (district level) was used to train and test the prediction algorithms, with 75\% of the data being the training set and the remaining 25\% being the test set. We consider a selection of the malarial indicators including the number of healthcare facilities, PfPR, reported malaria cases as a set of the independent variables, and malaria-attributed deaths as the dependent variable.

Since decision trees are known for fast training, we used the {\tt DecisionTreeRegressor} available in {\tt scikit-learn} package in Python with variable {\tt max\_depths} set to {\tt 10} to identify overfitting the data. However, due to large variations in the dataset, the decision tree becomes unstable.

Logistic Regression performs better than decision trees if the signal to noise ratio
is low. Therefore, we used the {\tt Logistic Regression} function with different solvers including {\tt lbfgs}, {\tt newton-cg}, and {\tt liblinear}, among which, {\tt lbfgs} produced results with better accuracy. The result thus obtained is shown in Table~\ref{tab2}. This model is being extended to include more indicators as part of the future work.

\begin{table}[h]
\begin{center} \small
    
\begin{tabular}{l c c}
\hline
District & Actual    & Predicted \\
\hline
Bikita & 2 & 2   \\
Bindura & 21 & 22   \\
Marondera & 5 & 8  \\
Bubi & 0 & 0   \\
Chikomba & 3 & 3   \\
Chiredzi & 16 & 19   \\
Hwedza & 2 & 1 \\
Masvingo & 2 & 3  \\
Mt. Darwin & 21 & 21 \\
Shurugwi & 0 & 1  \\
Rushinga & 6 & 5  \\
Beitbridge & 8 & 13  \\
Uzumba (UMP) & 5 & 5 \\
Mutasa & 15 & 19 \\
\hline
\vspace{2.5pt}
\end{tabular}
\caption{Actual and Predicted malaria deaths for the districts in Zimbabwe (2015).}
\label{tab2}
\end{center}
\vspace{-12pt}

\end{table}

\section{Conclusions}
Our preliminary experimental results confirm that not all of the considered indicators show a correlation to malaria cases and deaths at district, province and country level resolutions. For example, PfPR is highest in the Kariba and Binga districts, yet Chipinge province has the highest number of reported cases and deaths. Therefore, we can infer that there is a strong possibility that other parasites and/or vectors, environmental or social factors affect malaria incidence in the region. We are working on other spatio-temporal factors at a finer resolution, to explore more about cities and provinces with very high malaria incidence rate. We also intend to continue this research further by applying other machine learning techniques such as boosting and neural networks on these data to gain more insights on regions that can have a higher malaria incidence rate in the future.

\section*{Acknowledgments}
This work was partially supported by a Bill \& Melinda Gates Foundation Grand Challenges Explorations grant.